\documentclass{article}
\usepackage{fullpage}
\usepackage{graphicx}
\usepackage{hyperref}

\title{Against Fundamentalism}
\author{Matthew Leifer}
\date{October 11, 2018}

\begin{document}

\maketitle

\begin{abstract}
  In this essay, I argue that the idea that there is a most
  fundamental discipline, or level of reality, is mistaken.  My
  argument is a result of my experiences with the ``science wars'', a
  debate that raged between scientists and sociologists in the 1990's
  over whether science can lay claim to objective truth.  These
  debates shook my faith in \emph{physicalism}, i.e.\ the idea that
  everything boils down to physics.  I outline a theory of knowledge
  that I first proposed in my 2015 FQXi essay on which knowledge has
  the structure of a scale-free network.  In this theory, although
  some disciplines are in a sense ``more fundamental'' than others, we
  never get to a ``most fundamental'' discipline.  Instead, we get
  hubs of knowledge that have equal importance.  This structure can
  explain why many physicists believe that physics is fundamental,
  while some sociologists believe that sociology is fundamental.
  
  This updated version of the essay includes and appendix with my responses to the discussion of this essay on the FQXi website.
\end{abstract}

\section{What is Fundamental?}

As a physicist, it is easy to be impressed with the understanding that
fundamental physics has gifted us.  Through the ingenuity and hard
work of thousands of physicists, we have learned that all matter and
energy in the universe is composed of interacting quantum fields, and
we can in principle predict their behavior to great accuracy using the
standard model of particle physics.  On the large scale, Einstein's
theory of General Relativity, together with the standard model of
cosmology, give us an accurate picture of how the universe began, and
how it behaves on large scales.  Sure, there are a few phenomena that
are outside the scope of current physics, such as what happens in the
very early universe or near the singularity of a black hole, but, on
the scales relevant to human life, we have a pretty complete
understanding of all the relevant constituents of matter and
fundamental laws.  This picture is complete in the sense that it does
not seem to need any concepts from the other sciences, except perhaps
mathematics, in order to describe all matter.  In principle, we could
use fundamental physics to predict with the greatest possible accuracy
what will happen in any given situation, including those relevant to
chemistry and biology, and even in those sciences that deal with the
human mind, such as neuroscience, psychology, and sociology.  I say
``in principle'' because those calculations would involve an
impossibly detailed description of the initial conditions of the
system being studied, as well as infeasible computational power.  It
would be essentially impossible to identify and model a biological
system directly in terms of its constituent quantum fields.  So we can
admit that, in practice, biological explanations of how cells operate
are much more useful than descriptions in terms of fundamental
physics.  However, the question ``what is fundamental?''  concerns
what is possible in principle rather than with what is possible in
practice.

The view outlined above, that everything boils down to physics, is
called \emph{physicalism}.  Although it is an attractive view for a
physicist---I, personally, was drawn to physics because it seemed to
be the only way to truly understand the fundamental nature of
reality---I shall be arguing for precisely the opposite view in this
essay.  My position is deeply influenced by the ``Science Wars''; a
battle that raged in the 1990's between scientists, philosophers, and
sociologists over whether science can lay claim to objective truth.
In many ways, I am a casualty of the science wars, since they were at
their peak during my undergraduate education.  Being young enough not
to have developed strong opinions about the meaning of science, I have
been influenced by the sociology camp to a greater extent than most
scientists.  The extreme version of the sociology side of the
argument, which I call \emph{sociologism}, claims not only that
science is not objectively true, but that sociology is more
fundamental than physics.  It is quite understandable that a
sociologist might find this view as appealing as physicalism is to a
physicist, and a bit surprising that we do not have even more ``isms''
where scholars seek to put their own discipline at the top of the
tree.

Although I want to incorporate some of the sociological insights
into my argument, of course I view sociologism as just as barmy as
physicalism.  However, the fact that scholars can seriously argue that
a discipline other than physics should be considered fundamental lends
some support to my thesis that ``fundamental'' is a mistaken category.
If this is so, then we shall need a theory of knowledge that accounts
for the fact that subjects like physics can seem more ``fundamental''
than others when this is not actually so.  I shall attempt to develop
such a theory as well.

This essay is a sequel to my 2015 FQXi essay ``Mathematics is
Physics''\cite{Leifer2016}, in which I proposed a theory of knowledge
intended to explain why it is not surprising that advanced mathematics
is so useful in physics.  The theory of knowledge employed here is the
exact same one, but I want to relate it more explicitly to my thoughts
on the science wars.

\section{Dispatches from the Science Wars}

I would like to begin with the story of my first encounter with the
science wars.  When I was an undergraduate studying physics at
Manchester, my brother, who studied philosophy, got me interested in
philosophy of science by asking me difficult questions over dinner.
Since the well-being of my psyche depended on being able to defend the
position that physics is the most fundamental way of understanding the
world, I jumped at the chance to take a course entitled ``The Nature
of Scientific Enquiry'' when the opportunity came up.  At the very
least, I figured, it would help me win dinner table arguments with my
brother.

When the time came, I went to see my director of studies, the late
Dr.\ Anthony Phillips (still the best physics teacher I have ever
known) to tell him that I wanted to enrol in the course.  His first
reaction was, ``Wouldn't you rather take a course in fluid
mechanics?''.  After I rejected that option his next response was,
``OK, but don't believe a word they tell you.''  At the time, I
thought this rather uncollegial, but I did not realize that the course
was run by the sociology department, and was being taught by a
proponent of the ``strong program'' in the sociology of scientific
knowledge (SSK), a major school on the sociology side of the science
wars.  I did not know that we were supposed to be at war, but, in
light of that, Dr.\ Phillips comments make a lot more sense.

The first half of the course proceeded along the lines of a generic
philosophy of science course.  We studied Bacon \cite{Klein2016},
logical positivism \cite{Creath2017}, Popper \cite{Popper2001}, Kuhn
\cite{Kuhn1996}, and Lakatos \cite{Lakatos1970}.  However, unlike a
standard philosophy course, Kuhn was given a ringing endorsement, and
then we went off to study SSK.

The strong program of SSK is most closely associated with David Bloor
and his collaborators at the University of Edinburgh \cite{Bloor1991}.
It is intended as a response to earlier approaches to the sociology of
science, which are deemed ``weak''.  In ``weak'' studies, sociological
factors are only deemed important in understanding why ``failed'' or
``false'' theories are sometimes accepted.  For example, one might
look at how Stalin's totalitarian rule allowed Lysenco's ideas of
environmentally acquired inheritance to become the dominant theory of
genetics in the Soviet Union in the 1930's and 40's.  In modern times,
one might look at why the anti-vaccine movement or the idea that human
activity is not causing climate change are being increasingly accepted
in large segments of the US population.

In contrast to this, the strong program states that sociological
factors are equally important in understanding how successful
scientific theories, which are usually deemed ``true'', gain
acceptance.  If a theory is accepted science, it is very easy to fall
back on the argument that the reason it became accepted is simply that
it is ``true''.  Proponents of the strong program reject this
asymmetry of explanation, and want to study the sociological reasons
why science progresses the way it does period, without regard to
whether a or not a theory is ``true''.  In order to do this they
adopt, as a methodological principle, a ban on using the ``truth'' or
``correctness'' of a theory an an explanation for its acceptance.

Although this ban is supposed to be merely methodological---a
corrective for decades of studies which ignored sociological factors
other than in cases of ``error''---studies in the strong program tend
to show strong sociological influences in every case they look at.
Unless you are being deliberately contrarian, it is very hard not to
infer that, if you can actually find sociological reasons why theories
are accepted in every case, then scientific theories must be social
constructs, with no claim to be the ultimate arbiters of objective
truth.  Although defenders of the strong program like to emphasize
that the ban is meant to be methodological, and they are simply
``hands-off'' on the question of ultimate truth, it is pretty bizarre
to adhere to a methodology and, at the same time, not contemplate the
most obvious reason why that methodology might work well.  This leads
to cultural relativism about scientific truth and, despite protests to
the contrary, the language of cultural relativism does seep through
the rhetoric of the strong program.  Nonetheless, I define
``sociologism'' as the position that scientific theories are merely
social constructs, in contrast to the strong program itself, which
insists on only adopting this as a methodology.  Sociologism implies
that sociology is the most fundamental science, since it means that
understanding the content of any scientific theory is equivalent to
understanding the social factors that led to its acceptance.

To see how easily SSK devolves into sociologism, I want to relate an
experience from the Nature of Scientific Enquiry course.  In one of
our assignments, we were asked the question, ``If sociological factors
always play a role in determining which scientific theories are
accepted, does science still tell us anything about the real world?''
In the seminar discussion of this, the graduate TA proposed the
answer, ``Yes, because sociology is a science, so the study of
sociological factors is still a study of the real world.''  This is
sociologism writ large.  Not only do proponents of sociologism want to
take physicists down a peg or two, but they also want to view their
own subject as more fundamental than the sciences they are studying.
Everything hangs off sociology, as it were.

It is easy to ridicule sociologism.  After all, advocates of this view
still get on airplanes to fly to conferences.  If you really believe
that science is just a social construct, then you have no good reason
for believing the airplane will not simply fall out of the sky.  I,
for one, would not take the fact that flying airplanes is a tradition
of my culture as a convincing argument to get on board.  So,
proponents of this view seem to act like they believe at least some
aspects of science are objectively true, while simultaneously
propounding the opposite.

In the throes of intellectual enquiry, it is common to adopt overly
extreme views, which later have to be walked back.  This happens all
the time on the speculative end of theoretical physics, e.g.\ the
claim that the universe is literally a quantum computer
\cite{LLoyd2012}, or that all entangled systems are literally
wormholes \cite{Maldacena2013}, or that the universe is made of
mathematics \cite{Tegmark2014}.  So let's not hoist all of sociology
on the petard of their most extreme proponents, and instead look at
the evidence on which their claims are based.

Most studies in the mould of the strong program proceed along the
following lines.  We first consider the modes of enquiry that are
claimed to be the hallmarks of the scientific method, including
such things as induction, falsifiability, the role of crucial
experiments, skepticism of hypotheses that are not strongly supported
by evidence, rational choice between programs of research, etc.
Whichever of these (often conflicting) accounts of scientific enquiry
you subscribe to, the sociologists find that they are violated in
almost every case they look at, and identify sociological factors that
played a role in theory choice instead.

There is not space to delve into specific examples here, so I will
just mention Collins and Pinch's study of the role of the
Michaelson-Morely experiment in the acceptance of Einstein's
relativity \cite{Collins1993}, since that is of relevance to fundamental
physics.  In the usual story told to students, the Michaelson-Morely
experiment is a crucial experiment that led physicists to reject the
luminiferous ether, i.e.\ the idea that light waves must propagate in
some medium in the same way that you cannot have water waves without
there being some water to do the waving.  The ether was replaced by
Einstein's theory, which eliminates it.  Collins and Pinch show that
Michaelson-Morely experiments never produced conclusive evidence
against the ether, despite attempts spanning several years under
different experimental conditions.

Now, one might argue that the weight of experimental evidence for
relativity that has been acquired since then is justification for its
acceptance today, but still it was accepted long before any of this
was acquired.  One might also argue that Einstein's theoretical
explanation of the symmetry of Maxwell's equations is the real reason
why relativity was accepted, but this was not universally regarded as
compelling at the time.  Indeed, the controversy over this is the
reason why Einstein won the Nobel prize for his explanation of the
photoelectric effect rather than for relativity.  While it is a
stretch to conclude from this that relativity is just a social
construct, the process of its acceptance was rather less rational than
one might otherwise believe.  At the very least, the story we tell
about how relativity became accepted, which is part of the pedagogy of
relativity, is largely a social construct.

However, the problem with case studies like these is that
philosophical theories of science are not supposed to have the same
status as mathematical theories.  In the latter, if you find one
counter-example to a theorem then the theorem is false\footnote{Of
  course, we always have the option of changing the definitions to
  make the theorem true, if doing so leads to a more useful theory,
  and this often happens in mathematics \cite{Lakatos1976}.}.  Instead,
philosophical theories of science propose norms, which we should
strive to adhere to if we want to create reliable scientific
knowledge.  These norms include skepticism of hypotheses that have no
evidential support, designing experiments that remove as much bias as
possible, etc.  Nobody is claiming that these norms are strictly
adhered to 100\% of the time, and that sociological factors play
absolutely no role.  Instead, the claim is that by attempting to
adhere to these norms, the community as a whole, over long periods of
time, will develop knowledge that is more reflective of the objective
world than otherwise.

To put it another way, the ``scientific method'' cannot really be
characterized in a precise way that is applicable to all cases.  For
any methodological principle that you might propose, one can find
cases where it is not really applicable.  But that does not mean that,
upon looking at the particulars of a specific theory, one cannot
decide whether the evidence supports it.  We may use different methods
and standards of evidence in fundamental physics, climate science, and
psychology, but these all bear a family resemblance, and an expert in
one of those fields can use the available evidence to decide how
likely a given claim is to be true.  The fact that we cannot give a
discipline-invariant definition of \emph{the} scientific method does
not seem to have gotten in the way of the progress of any scientific
discipline in particular.

Nonetheless, the studies of the strong program do show that social
factors have played a larger role in the construction of ``true''
theories than you might otherwise have thought, so the idea that we
should only pay attention to sociology in cases of ``error'' is
suspect.  Generally, all scientific discourse takes place within a
language, and is conducted by entities that are situated within
a society, with all the baggage that that entails, so social values are
implicitly used in the construction of science whether we like it or
not.  Although physics makes heavy use of mathematics, so is arguably
less influenced by the particulars of common language than other
sciences, few physicists believe that the content of physics is
entirely contained in its mathematical equations.  We need discourse
to understand what our theories mean, how they are connected to
observations, and even what questions are sensible to ask of the them.
Hence, the idea that physical theories may not be completely
objective, and that sociological factors may play a role in their very
construction, should at least be an option on the table, regardless of
how small or large you think that role is.

One example where sociological factors have had a strong influence on
physics is the dominance of the Copenhagen interpretation in the
foundations of quantum mechanics.  To modern eyes, it looks like the
founders of quantum mechanics jumped to conclusions about the nature
of (un)reality based on scant evidence.  While much evidence that can
be construed as supporting this kind of view has been acquired in the
meantime, the Copenhagen view was accepted by the majority
of physicists for decades without many physicists actually feeling the
need acquire this evidence.  Although there is more tolerance for
diverse views on the interpretation of quantum mechanics today,
Copenhagen has had a lasting influence on what physicists think a
physical theory should look like, which may be cutting off fruitful
research directions.

On the other hand, we do not want to endorse sociologism, in which we
cannot explain why airplanes do not fall out of the sky, why children
should be vaccinated, and why we should take action on climate change.
The success of our fundamental physical theories surely means
something for the objective physical world.  Therefore, we should not
replace the claim that physics is fundamental with the claim that
sociology is fundamental instead.  What we need is a theory
of knowledge that can account for why we should trust that airplanes
will not just fall out of the sky, but also allows external factors to
influence physics in a controlled way.  If it can also explain why
smart people can be led to believe that physics is fundamental, and
other smart people that sociology is fundamental, then so much the
better.

\section{A theory of knowledge}

To begin, I want to recall my own answer to the question of whether
science tells us anything about the real world, that I gave in my
undergraduate assignment.  It already contains the seeds of the more
sophisticated account I want to develop here.

Clearly, I reasoned, it is impossible that scientific theories have
nothing to do with the observed empirical world.  If a theory implied
that airplanes must necessarily always fall out of the sky, then we
would rightly reject such a theory as incorrect.  At any given time,
there is a large space of possible theories that are not in bald
conflict with the available empirical evidence.  When new evidence is
acquired, the size of that space is reduced.  It is still very large,
so sociological factors can play a strong role in determining which of
the theories in that space is ``true'', but the chosen theory still
tells us something about the objective physical world because we
cannot choose choose just any theory we like.  There is a constrained
surface of theories that are compatible with the evidence, and that
constraint is reflective of reality.

Whilst I think this is a reasonable response to the assignment
question, it is far from giving an accurate account of the nature of
knowledge.  This is because the set of theories that are compatible
with the evidence is still truly vast, and contains many things that
we would not want to call science.  For example, the theory that is
identical to current physics, but also posits that there are green
aliens hiding on the dark side of the moon that are completely
undetectable because they do not interact in any way with ordinary
matter, is compatible with current evidence, but we would not want to
call it scientific.  In the philosophical literature, this problem is
known as the underdetermination of theory by evidence.  This problem
does not seem to arise all that much in practice, so there are clearly
other constraints that determine what counts as knowledge.  Some of
these may come from social factors, and some from more objective
norms.  To resolve this, we have to look at the actual structure of
human knowledge.

Note that here I am diverging from what epistemologists (philosophers
who study the nature of knowledge) usually mean by a theory of
knowledge.  An epistemologist would usually define knowledge as
something like ``justified, true belief'' and study the way in which
knowledge is discovered as a separate question from whether it is
justified.  For example, if I have an intuition in the shower that
leads to a new theory of physics then I do not need to think about why
I came up with that intuition (the context of discovery) to understand
whether we should believe the theory (the context of justification).
I reject the distinction between the contexts of discovery and
justification because I think that key aspects of the process by which
we uncover new knowledge determine its relationship to other knowledge
and to the empirical world.

\begin{figure}[!htb]
  \centering
  \includegraphics[scale=0.7]{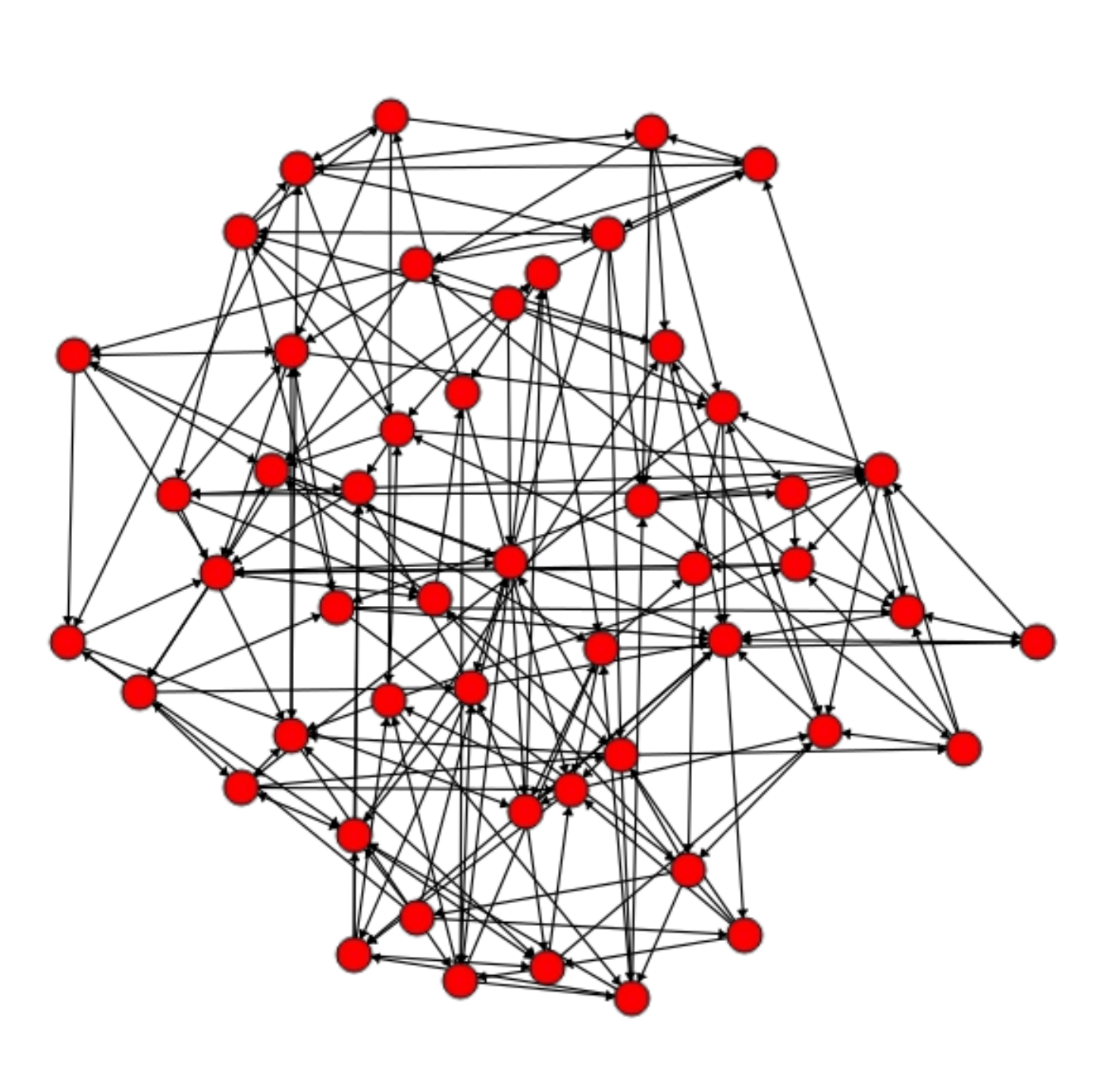}
  \caption{\label{network}Example of a network of nodes and links.}
\end{figure}

To understand the structure of knowledge, consider a network of nodes
connected by links (see fig.~\ref{network}).  The nodes are supposed to
represent items of knowledge.  These can include basic facts of
experience, e.g.\ ``that car looks red'', more abstract physical
facts, e.g.\ ``the charge of the electron is
$1.602 \times 10^{-19}\,\mbox{C}$.'', or even whole theories, e.g.\
``Electrodynamics''.  Clearly, the more abstract nodes can be broken
down into smaller constituents, e.g.\ we can break electrodynamics
down into its individual equations and explanations, so we can look at
the network at a higher or lower degree of abstraction or
coarse-graining.  The links represent a connection between items of
knowledge.  I do not want to be too specific about the nature of this
connection.  It could mean, ``can be derived from'', ``is a special
case of'', or even ``there is a strong analogy between''.  Depending
on the nature of the allowed connections, we would obtain slightly
different networks, but that is fine so long as we allow sufficient
types of connections to capture what we want to think of as the
structure of knowledge.

There is evidence that the knowledge network, so constructed, would
have the structure of a \emph{scale free network} \cite{Albert2002}.
Without getting into the formal definition of such networks, the
distribution of nodes and links in such networks has two important
properties.  Firstly, there are some nodes, called \emph{hubs}, which
have significantly more connections to other nodes than a typical
node.  Secondly, the shortest path you can take between two nodes by
following links is much shorter than you would think, given the total
number of nodes.  This second phenomenon is called ``six degrees of
separation'' after the idea that any two people on Earth can be
connected by friend-of-a-friend relationships in about six steps,
despite the fact that there are billions of people on Earth.

Now, obviously, I do not literally have the knowledge network to hand,
but there are real world networks that ought to approximate its
structure.  We could, for example, look at the structure of the world
wide web, where web pages are the nodes and hyperlinks are the links,
or do the same thing for Wikipedia articles.  We could take the nodes
to be scientific papers and draw a link when one paper cites another.
All of these examples have been found to approximate the structure of
a scale-free network \cite{Albert2002}.  Now, obviously, such networks
include things that we would not ordinarily want to call
``knowledge'', such as the name of Kanye West and Kim Kardashian's
latest baby, or authors citing their own papers for no other reason
than to increase their citation count.  However, whenever a society of
intelligent agents form a network of connections organically by a
large number of individual actions, they seem to do so in a scale-free
way.  Since the knowledge network is generated in this way, it seems
likely that it would be scale-free too.

In my 2015 FQXi essay, I gave a mechanism for the generation of
knowledge by abstraction from analogies that could plausibly lead to a
scale-free knowledge network.  This process starts with nodes that
represent the blooming, buzzing confusion of raw experience, which
will end up being the nodes at the edges of the networks.  We then
draw analogies between nodes that are similar and, at some point,
develop a higher level abstraction to capture the commonalities of
those nodes.  The links between every analogous node are then replaced
with links to the higher level node, which reduces the number of
links and complexity of the network.  This process continues at higher
and higher levels of abstraction, drawing analogies between higher
level nodes and then replacing those by further abstractions.  For
further details, I refer to my 2015 essay.

Here, I want to make a few points about the structure of the network
so generated.  Firstly, the ``real world'' imposes itself on the
network by the edge nodes that represent raw experience.  The
commonalities of those nodes impose the set of analogies it is
possible to draw, and hence the abstractions it is meaningful to
define.  In this way, the empirical world imposes itself on even very
high level abstractions, such as the fundamental physical theories, so
those theories do reflect the structure of the physical world.
However, there are also many ways in which societal contingencies
affect the structure of the network, e.g.\ the interests of the
participating agents affect the order in which analogies and
abstractions are drawn, which can affect the global structure of the
network.  So we can have a strong role for both the physical world and
sociological factors in determining what we regard as the ``true''
structure of knowledge.

It is important to note that \emph{any} large set of interacting
agents attempting to make sense of the world could use this process to
generate a scale-free knowledge network.  Intelligent aliens or
artificial intelligences would work just as well as humans.  What is
important is that there are independent entities interacting via
social connections.  The structure of the network is partly reflective
of the structure of the world, and partly reflective of the fact that
a social network of agents is generating the knowledge.  I do not
really think that it makes sense to speak of ``knowledge'' outside
this context.  For me, knowledge is necessarily a shared
understanding.

At this point, one might ask why a scale-free network is a good way of
organizing knowledge, i.e.\ why would nature endow us with the
capability to organize knowledge in this way?  Any given agent can
only learn a small part of the knowledge network.  The hub nodes
encode a lot of information at a high level of abstraction, such that
it is possible to get to any other node in a relatively short number
of steps.  Our fundamental theories of physics, as well as general
theories of sociology, are examples of such hub nodes.  In our
undergraduate studies, we tend to learn a lot about a single hub node,
and work outwards from that as we increase our specialization.  The
existence of hubs ensures that the six degrees of separation property
holds, so that it is possible to get from any two specialized
disciplines to a common ground of knowledge in a relatively short
number of steps.  If, for example, we encounter a problem that
requires both a physicist and a biologist to solve, they can work back
to a hub that both of them understand and use that as their starting
point.  This enables efficient collaboration between disciplines.  In
general, scale-free networks are a very efficient way of encoding
information.

The scale-free structure also explains why smart physicists can think
that physics is fundamental, while similarly smart sociologists can
think sociology is fundamental.  If you only learn a limited number of
nodes hanging off a single hub nodes, then the structure of your
knowledge is hierarchical, with everything seeming to hang off the
hub.  If you are a physicist, with fundamental physics as your hub,
you will see physics as fundamental to everything, whereas if you have
a sociological hub you will see sociology everywhere.  The reality is
that there are several hubs, all with equal importance, that abstract
different aspects of human experience.  Both physicalism and
sociologism assume a hierarchical structure of knowledge, with a
different discipline at the top.  If, in fact, the structure of
knowledge is not a hierarchy, then the question of which discipline is
the most fundamental simply evaporates.  Now, of course, hub nodes are
more important than other nodes because they encode a larger portion
of human knowledge, so it does make sense to think of them as more
fundamental than the other nodes, but there is no sense in which
everything boils down to a single most fundamental node.

\section{Conclusion}

In conclusion, if human knowledge has the structure of a scale-free
network, which is as much a feature of the fact that it is generated
by a society of interacting agents as it is reflective of the physical
world, then there is no sense in talking about a most fundamental area
of knowledge.  The question, ``what is fundamental?'' simply
evaporates.

Although I have argued that physical knowledge is reflective of
physical reality, we still have the question of how objective it is.
Does the physics knowledge network necessarily have to look similar to
our current theories of physics, or could there be a very dissimilar
looking network that is equally efficient, formed on the basis of
the same evidence?  Even if we think of the process of acquiring
knowledge as looking for the most efficient scale-free encoding, there
could be local minima in the space of all possible networks, which
would be difficult for a process based on locally adding nodes and
replacing links to get out of.  If two societies can end up with very
different networks based on the same process, then this lends weight to
the argument that social construction is the dominant influence of
scientific theories.  However, if the physics networks generated by
this process all tend to look the same up to minor differences, then
they are more reflective of the world than of society.

I view this as an empirical question.  If we ever encounter an
advanced alien civilization that has developed in isolation from us, will
its physics network look similar to ours or not?  I think it is likely
that the answer is yes, but that is not something I can prove.
Barring contact with aliens, we could answer the same question by
placing a network of sufficiently advanced artificial intelligences on
a knowledge gathering quest.  This is obviously not a question we can
answer right now, but maybe one day we will.

\appendix

\section{Responses to Online Discussion}

Since it was posted on the FQXi wesbite, this essay has generated an interesting online discussion.  Unfortunately, I was not able to participate actively in the discussion at the time, so I respond to some of the more interesting comments here.  There is not space to address every comment, so interested readers are encouraged to read the full discussion online \cite{Various2018}.

Jochen Szangolies argues that the reliable convergence of ideas in physics should be taken as evidence that physics is objective and fundamental, citing the historical example of the convergence of measurements of the charge to mass ratio of the electron.  However, an advocate of sociologism could equally argue that sociological factors are responsible for the convergence.  There would be sociological pressure to come up with a unique theory.  Discussions of which methods of approximation are appropriate, which systematic errors to take into account, which methods of measurement are most accurate, and which methods of data analysis to use, all occur within the scientific community.  These are primarily responsible for convergence, and could be affected by sociological factors.  Of course, I do not personally believe that sociological factors are primary in this process, but convergence of ideas in physics is not the knockdown argument against sociologism that it might appear to be.

Szangolies also argues that there is some ambiguity over what constitutes a ``node'' and what constitutes an ``edge'' in the knowledge network.  He cites the example that if ``Socrates is a man'' and ``Socrates is mortal'' are nodes, then the derivation of the latter from the former is connected by the edge ``All men are mortal'', which could also be construed as an item of knowledge, and hence a node.  Note that we could look at this example differently, viewing all three items as nodes, and the rules of categorical syllogism as the connecting edge, but then perhaps these rules should themselves be a knowledge node.

I was deliberately vague about what should constitute a node and what should constitute an edge in the essay, precisely because of this sort of ambiguity.  The network can be constructed at various levels of coarse-graining, depending on what we want to regard as the units of knowledge, e.g.\ scientific papers, entire theories, basic facts, etc.  However, scale-free networks are self-similar, which means that the coarse-graining of such a network would also be scale-free, so to a large degree it should not matter exactly how we construct it.  It is also important to realize that the knowledge network is only a model for the structure of knowledge, that I hope caputes important features of that structure, but cannot be expected to capture all subtleties.  In this sense, it is like a model in physics, where carefully chosen approximations are made in order to yield a useful explanatory theory because working directly with the fundamental equations would be too complicated.  I am open to the idea that a more general discrete combinatorial structure might better represent the structure of knowledge, e.g.\ a hypergraph in which more than two nodes can be linked by a hyperedge.  The only important thing is that we can define a notion of scale-free for that structure and that a network can be used to approximate it.  The network structure of the scientific citation network, the world wide web, and Wikipedia are meant to serve as evidence that knowledge can be approximately represented this way, but I freely admit that there are subtleties in the structure of knowledge that are not fully captured by these models.

Szangolies also points out that my knowledge network is epistemic, and does not deal with the ontic structure of the world, i.e.\ what is really out there.  I acknowledge that this criticism is appropriate from a scientific realist point of view, but I adhere much more closely to a pragmatist theory of truth, in which what is true rougly corresponds to what is ``useful''.  This means I view my epistemic account of knowledge as more fundamental than any ontic account, and am skeptical about the meaning of the latter.  I am committed to a naturalist metaphysics, in the sense that I think we must look at how the things we call knowledge are actually acquired, rather than positing an a priori structure that they must fit into.

John C. Hodges points out that human societies have often adopted similar social structures, and that Darwinian natural selection may be responsible for this.  A scale-free network is an efficient way of encoding knowledge, and I agree that once evolution has produced an intelligent social species, there would be Darwinian pressure to structure society in this way.  So I expect alien species to structure their knowledge in a scale-free network, but this still leaves open the question of whether there is more than one local minimum for the structure of a knowledge network representing our universe.
 
Ken Wharton argues that the structure of a knowledge network can still be used to assert that physics is fundamental, in the sense that, as a hub node, it is more fundamental than non-hub nodes.  Indeed, I recognize that the question of ``more fundamental'' makes sense.  What I reject is the notion of ``most fundamental'' and the idea, common among physicists, that physics has the special status of being more fundamental than anything else.

Cristinel Stoica posits the idea that, since the world is fundamentally quantum mechanical, the knowledge network should be viewed as emergent from a unitarily evolving quantum state of the universe.  Since I am not a straightforward realist about our scientific theories, I strongly reject this idea.  The structure of the knowledge network determines in part the structure of our scientific theories, so I would say that quantum states are emergent from the network rather than the other way round.

Alyssa Ney points out the similarity between my view of knowledge and that posted by Quine in his essay, ``Two Dogmas of Empiricism'' \cite{Quine1951}.  Indeed, Quine is a major influence on my thinking, and I thank Ney for giving me a reason to reread this essay.  Quine writes:
\begin{quotation}
The totality of our so-called knowledge or beliefs, from the most casual matters of geography and history to the profoundest laws of atomic physics or even of pure mathematics and logic, is a man-made fabric which impinges on experience only along the edges. Or, to change the figure, total science is like a field of force whose boundary conditions are experience. --- W.\ V.\ Quine \cite{Quine1951}.
\end{quotation}
This is quite similar to my view of the importance of realizing that knowledge is constructed by societies and the role of experience at the edges of our knowledge network.

Ney also questions whether physicalism is in conflict with the strong program in the sociology of science.  She argues that even if we have sociological explanations for the uptake of physical theories over time, this does not rule out the idea that there is also a more fundamental physical explanation for why they are true.

While this is true of the formal definition of the strong program, in which the use of the truth of a scientific theory as an explanation for its acceptance is rejected as a methodological principle, I believe that most advocates of this program are (at least covertly) social constructivists.  Indeed, if you find sociological reasons for the uptake of physical theories everywhere you look then it becomes difficult to believe that any other explanation for their success is needed, and a descent into sociologism is likely, if not inevitable.  Even rejecting sociologism, from my point of view, which is more pragmatist rather than realist, I find it difficult to understand what a ``physical explanation'' would actually mean in this context.  Once I have explained why the theory is a useful addition to the knowledge network, in the sense of enabling an efficient encoding of experience in a scale-free way, I do not see what else is left to explain.  I acknowledge that this account is not complete according to scientific realism, but debating the relative merits or realism and pragmatism will have to wait for a future essay contest. 

\bibliographystyle{unsrturl}
\bibliography{FQXi2017}

\begin{thebibliography}{10}

\bibitem{Leifer2016}
M.~S. Leifer.
\newblock Mathematics is physics.
\newblock In A.~Aguirre, B.~Foster, and A.~Merali, editors, {\em Trick or
  Truth? The Mysterious Connection Between Physics and Mathematics}, pages
  21--40. Springer, 2016.

\bibitem{Klein2016}
J.~Klein.
\newblock Francis bacon.
\newblock In E.~N. Zalta, editor, {\em The Stanford Encyclopedia of
  Philosophy}. Winter 2016 edition, 2016.
\newblock URL:
  \url{https://plato.stanford.edu/archives/win2016/entries/francis-bacon/}.

\bibitem{Creath2017}
R.~Creath.
\newblock Logical empiricism.
\newblock In E.~N. Zalta, editor, {\em The Stanford Encyclopedia of
  Philosophy}. Fall 2017 edition, 2017.
\newblock URL:
  \url{https://plato.stanford.edu/archives/fall2017/entries/logical-empiricism/}.

\bibitem{Popper2001}
K.~Popper.
\newblock {\em The Logic of Scientific Discovery}.
\newblock Routledge Classics, 2001.

\bibitem{Kuhn1996}
T.~S. Kuhn.
\newblock {\em The Structure of Scientific Revolutions}.
\newblock The University of Chicago Press, third edition, 1996.

\bibitem{Lakatos1970}
I.~Lakatos and A.~Musgrave, editors.
\newblock {\em Criticism and the Growth of Knowledge}.
\newblock Cambridge University Press, 1970.

\bibitem{Bloor1991}
D.~Bloor.
\newblock {\em Knowledge and Social Imagery}.
\newblock University of Chicago Press, second edition, 1991.

\bibitem{LLoyd2012}
S.~Lloyd.
\newblock The universe as a quantum computer.
\newblock In H.~Zenil, editor, {\em A Computable Universe: Understanding and
  Exploring Nature as Computation}, pages 569--584. Wold Scientific, 2012.

\bibitem{Maldacena2013}
J.~Maldacena and L.~Susskind.
\newblock Cool horizons for entangled black holes.
\newblock {\em Fortschritte der Physik}, 61(9):781--811, 2013.

\bibitem{Tegmark2014}
M.~Tegmark.
\newblock {\em Our Mathematical Universe: My Quest for the Ultimate Nature of
  Reality}.
\newblock Alfred A. Knopf, 2014.

\bibitem{Collins1993}
H.~Collins and T.~Pinch.
\newblock {\em The Golem: What You Should Know About Science}.
\newblock Cambridge University Press, 1993.

\bibitem{Lakatos1976}
I.~Lakatos.
\newblock {\em Proofs and Refutations: The Logic of Mathematical Discovery}.
\newblock Cambridge University Press, 1976.

\bibitem{Albert2002}
Réka Albert and Albert-László Barabási.
\newblock Statistical mechanics of complex networks.
\newblock {\em Reviews of Modern Physics}, 74(1):47--96, 2002.
\newblock eprint arXiv:cond-mat/0106096.
\newblock \href {http://dx.doi.org/10.1103/RevModPhys.74.47}
  {\path{doi:10.1103/RevModPhys.74.47}}.

\bibitem{Various2018}
Various.
\newblock Online discussion of ``against fundamentalism" by matthew saul
  leifer.
\newblock \url{https://fqxi.org/community/forum/topic/3111}, 2018.

\bibitem{Quine1951}
W.~V. Quine.
\newblock Two dogmas of empiricism.
\newblock {\em The Philosophical Review}, 60(1):20--43, 1951.
\newblock \href {http://dx.doi.org/10.2307/2181906}
  {\path{doi:10.2307/2181906}}.

\end{thebibliography}

\end{document}